\documentclass[12.5pt]{article}
\usepackage{graphicx}
\begin{document}

\twocolumn[\begin{center}
{\large {\bf Waves in General Relativistic Two-fluid Plasma\\ around a Schwarzschild Black Hole}}\\
\vspace{.4cm}
M. Atiqur Rahman\\
{\it Department of Applied Mathematics, Rajshahi University,
Rajshahi - 6205, Bangladesh}
\end{center}
\vspace{.4cm}
\centerline{\bf Abstract}
\baselineskip=18pt
\begin{center}
\parbox{15cm}{Waves propagating in the relativistic electron-positron or ions plasma are investigated in a frame of two-fluid equations using the $3+1$ formalism of general relativity developed by Thorne, Price and Macdonald (TPM). The plasma is assumed to be freefalling in the radial direction toward the event horizon due to the strong gravitational field of a Schwarzschild black hole. The local dispersion relations for transverse and longitudinal waves have been derived, in analogy with the special relativistic formulation as explained in an earlier paper, to take account of relativistic effects due to the event horizon using WKB approximation.
{\it PACS}: 95.30.Qd, 95.30.Sf, 97.60.Lf\\
{E-mail: $atirubd@yahoo.com$}}
\end{center}

]{\bf \hspace{1.5 cm} \large 1. Introduction}\label {sec1}\\
In recent year plasma equations and general relativity are usually considered together. The Coulomb potential of charge particles due to coupling is much stronger than the gravitational potential and are often neglected in the Newtonian approximation. But the mean gravitational field for certain astronomical objects like galactic nuclei or black holes may be strong and the observation of magnetic fields indicates that a combination of general relativity and plasma physics at least on the level of a fluid description is appropriate. The plasma in the black hole environment may act as a fluid and black holes greatly affect the surrounding plasma medium (which is highly magnetized) with their enormous gravitational fields. Hence plasma physics in the vicinity of a black hole has become a subject of great interest in astrophysics. It is therefore of interest to formulate plasma physics problems in the context of general relativity.

Thorne and MacDonald \cite{one,two} have introduced Maxwell\rq s equations in 3+1 coordinates, which provides a foundation for formulation of a general relativistic (GR) set of plasma physics equations in the strong gravitational field of both the nonrotating and rotating black holes and the \lq\lq membrane paradigm\rq\rq \cite{three} is a good example of such a formalism in which the electromagnetic equations and the plasma physics at least look somewhat similar to the usual formulations in flat spacetime while taking accurate account of general relativistic effects such as curvature. The membrane paradigm is mathematically equivalent to the standard, full general relativistic theory of black holes, so far as all physics outside the horizon is taken into account.

Sakai and Kawata (SK) \cite{four} have developed the linearized treatment of plasma waves using special relativistic formulation. Such an investigation of wave propagation in a general relativistic two-fluid plasmas near a black hole is important for an understanding of plasma processes. That is, what happens when the plasma are assumed to be freefalling onto the black hole. The study of plasma wave in the presence of strong gravitational fields using the $3+1$ approach is still in its early stages. Zhang \cite{five,six} has considered the care of ideal magneto hydrodynamics waves near a Kerr black hole, accreting for the effects of the holes angular momentum but ignoring the effects due to the black hole horizon. Holcomb and Tajima \cite{seven}, Holcomb \cite{eight}, and Dettmann et. al. \cite{nine} have considered some properties of wave propagation in a Friedmann universe. Daniel and Tajima \cite{ten} studied the physics of high frequency electromagnetic waves in a strong Schwarzschild plasma. Marklund et al \cite{eleven} have found a mode representing high frequency plasma oscillation in a charged two-component plasma using the exact 1+3 covariant dynamical fluid equations in the presence of electromagnetic fields about a Friedmann-Robertson-Walker model by ignoring the fluid\rq s thermal effects.  Servin et al \cite{twelve} and Kleidis et al \cite{thirteen,fourteen} have studied the propagation of gravitational waves in a collisionless plasma with an external magnetic field parallel to the direction of propagation, while Forsberg et al \cite{fifteen} have presented an investigation of nonlinear interactions between gravitational radiation and modified Alfv\'{e}n modes in astrophysical dusty plasmas.

There is also work on fluid dynamics and kinetic gas theory in the context of cosmology. The book by Bernstein \cite{sixteen} treats gas kinetics in the Friedmann-Lema\^{\i}tre-Robertson-Walker (FLRW) model. However, there are relatively few relativistic cosmological investigations that take into account plasma effects and the behavior of matter in the presence of electromagnetic fields \cite{seventeen,eighteen,nineteen,twenty,twenty one,twenty two,twenty three}. Therefore, the general relativistic treatment of plasmas, both in astrophysics as well as in plasma physics, seems to be a field open to investigation.

A plasma can propagate both linear and nonlinear waves. Linear refers to the simplifying approximations that are possible for small amplitude waves like, Alv\'en and high frequency electromagnetic waves, and nonlinear refers to large amplitude phenomena not predicted by linear models. In this paper, the set of two-fluid equations for collisionless ideal plasma are used to make an initial attempt to be the discovery of an instability caused by the general relativistic term in the dispersion relations for transverse (electromagnetic) and longitudinal (electrostatic) waves using action principle for a hot plasma developed by Heintzmann and Novello \cite{twenty four}. Similar multifluid equations have recently been used to calculate local dispersion laws for plasma waves in strong and weak gravitational fields; see, e.g., Buzzi et al \cite{twenty five, twenty six}, and Rahman et al \cite{twenty seven, twenty eight, twenty nine, thirty}. Thus, the present work will form the essential basis of nonlinear, more complicated, investigations.

In the present paper Sec. 2 summarize the 3+1 formulation of general relativity. In Sec. 3 we review the two-fluid plasmas governing equations in Schwarzschild coordinates. The transverse and longitudinal parts are separated by introducing a new complex transverse fields and velocities using Rindler coordinates in Sec. 4. In Sec. 5  the two fluid equations are linearized for wave propagation by giving a small perturbation to fields and fluid parameters. We discuss the way in which the unperturbed fields and fluid parameters and their derivatives with respect to $z$ depend on the surface gravity of the black hole and freefall velocity in Sec. 6. In Sec. 7 the two-fluid equations are simplified using action principle developed by Heintzmann and Novello \cite{twenty four}. The dispersion relations for transverse and longitudinal waves are developed and solved using the analytical method developed by Mikhailovskii \cite{thirty one} in Sec. 8 and 9.  Finally,  we present our remarks in Sec. 10. Here, we use units in which $G=c=k_B=1$.\\

{\bf \large 2. 3+1 Formalism of Schwarzschild Spacetime}\label {sec2}\\
Our work presented in this paper is based on the 3+1 formulation of general relativity developed by Thorne, Price, and Macdonald (TPM) \cite{one,two,three}. The basic concept behind the $3+1$ formulation of general relativity is to select a preferred set of spacelike hypersurfaces which form the level surfaces of a congruence of timelike curves. A particular set of these hypersurfaces constitutes a time slicing of spacetime. The hypersurfaces considered here are of constant universal time $t$. In the $3+1$ formulation, the Schwarzschild metric is given by
\begin{eqnarray}
ds^2=-\alpha^2dt^2+\frac{1}{\alpha^2}dr^2+r^2(d\theta ^2+{\rm sin}^2\theta d\varphi ^2),\label{eq1}
\end{eqnarray}
where the lapse function $\alpha$ is given by
\begin{equation}
\alpha =\sqrt{1-{2M}/{r}}.\label{eq2}
\end{equation}
The hypersurfaces of constant universal time $ t$ define an absolute three-dimensional space described by the metric
\begin{eqnarray}
ds^2=\frac{1}{\alpha^2}dr^2+r^2(d\theta ^2+{\rm sin}^2\theta d\varphi ^2).\label{eq3}
\end{eqnarray}

We consider a set of fiducial observers (FIDOs) having properties as mentioned by Buzzi et al \cite{twenty five} remaining at rest with respect to this absolute space. For a detailed concept of a set of fiducial observer (FIDOs) see the membrane paradigm book \cite{three}. To make the local measurements of all physical quantities FIDOs use a local Cartesian coordinate system that have basis vectors of unit length tangent to the coordinate lines:
\begin{equation}
{\bf e}_{\hat r}=\alpha \frac{\partial }{\partial r},\hspace{.2cm}{\bf e}_{\hat \theta }=\frac{1}{r}\frac{\partial }{\partial \theta },\hspace{.2cm}{\bf e}_{\hat \varphi }=\frac{1}{r{\rm sin}\theta }\frac{\partial }{\partial \varphi}\label{eq4}.
\end{equation}
For a spacetime viewpoint rather than a $3+1$ split of spacetime, the set of orthonormal vectors also includes the basis vector for the time coordinate is given by
\begin{equation}
{\bf e}_{\hat 0}=\frac{d}{d\tau }=\frac{1}{\alpha }\frac{\partial }{\partial t}\label{eq5},
\end{equation}
The lapse function $\alpha$ plays the role of a gravitational potential and thereby governs the ticking rates of clocks and redshifts. We can calculate the gravitational acceleration felt by a FIDO from the lapse function as follows \cite{one,two,three}:
\begin{equation}
{\bf a}=-\nabla {\rm ln}\alpha =-\frac{1}{\alpha }{\frac{M}{r^2}}{\bf e}_{\hat r}\label{eq6},
\end{equation}
Equation (\ref{eq6}) shows that far from the black hole event horizon the gravitational acceleration become weak and approaches the Newtonian value for flat spacetime. However, near the horizon, the gravitational acceleration approaches infinity as $\alpha \rightarrow 0$. The rate of change of any scalar physical quantity or any three-dimensional vector or tensor, as measured by a FIDO, is defined by the derivative
\begin{equation}
\frac{D}{D\tau }\equiv \left(\frac{1}{\alpha }\frac{\partial }{\partial t}+{\bf v}\cdot \nabla \right),\label{eq7}
\end{equation}
$\bf v$ being the velocity of a fluid as measured locally by a FIDO. Since all the quantities are measured locally by the FIDO, all the vector quantities are neither covariant nor contravariant.\\

{\bf \hspace{1.2 cm} \large 3. Two-fluid Equations}\label {sec3}\\
We consider two-component plasma such as an electron-positron or electron-ion. The basic equations (the equation of continuity and Maxwell\rq s equations) for each fluid of species $s$ with velocity ${\bf v}_s$, number density $n_s$, and the relativistic Lorentz factor $ \gamma _s$ as derived by TPM \cite{one,two,three} and Buzzi et al \cite{twenty five, twenty six} using the 3+1 formulation are given by
\begin{equation}
\frac{\partial }{\partial t}(\gamma _sn_s)+\nabla \cdot (\alpha \gamma _sn_s{\bf v}_s)=0\label{eq8},
\end{equation}
and
\begin{eqnarray}
\nabla \cdot {\bf B}&=&0,\label{eq9}\\
\nabla \cdot {\bf E}&=&4\pi \sigma ,\label{eq10}\\
\frac{\partial {\bf B}}{\partial t}&=&-\nabla \times (\alpha {\bf E}),\label{eq11}\\
\frac{\partial {\bf E}}{\partial t}&=&\nabla \times (\alpha {\bf B})-4\pi \alpha {\bf J},\label{eq12}
\end{eqnarray}
where the charge and current densities are defined with charge $q_s$ of each fluid species by
\begin{equation}
\sigma =\sum_s\gamma _sq_sn_s,\hspace{1.2cm}{\bf J}=\sum_s\gamma _sq_sn_s{\bf v}_s\label{eq13},
\end{equation}
where $s$ is $1$ for electrons and $2$ for positrons (or ions). The presence of lapse function $\alpha$ signifies the general relativistic effect around a Schwarzschild black hole.

Using Maxwell's Eqs. (\ref{eq11}) and (\ref{eq12}) the equations for the conservation of energy and momentum, as derived by TPM \cite{one,two,three} and Buzzi et al \cite{twenty five, twenty six}, for Schwarzschild black hole are given by
\begin{eqnarray}
\frac{1}{\alpha }\frac{\partial }{\partial t}\epsilon _s&=&-\nabla \cdot {\bf S}_s+2{\bf a}\cdot {\bf S}_s,\label{eq14}\\
\frac{1}{\alpha }\frac{\partial }{\partial t}{\bf S}_s&=&\epsilon _s{\bf a}-\frac{1}{\alpha }\nabla \cdot (\alpha {\stackrel{\longleftrightarrow }{\bf W}}_s)\label{eq15},
\end{eqnarray}
where the energy density $\epsilon _s$, the momentum density ${\bf S}_s$, and stress-energy tensor $W_s^{jk}$ for the electromagnetic field as
\begin{eqnarray}
\epsilon _s=\frac{1}{8\pi }({\bf E}^2+{\bf B}^2),\quad{\bf S}_s=\frac{1}{4\pi }{\bf E}\times {\bf B},\nonumber\\
W_s^{jk}=\frac{1}{8\pi }({\bf E}^2+{\bf B}^2)g^{jk}-\frac{1}{4\pi }(E^jE^k+B^jB^k)\label{eq16}.
\end{eqnarray}
For a perfect relativistic fluid of species $s$ in three-dimensions, the energy density $\epsilon _s$, the momentum density ${\bf S}_s$, and stress-energy tensor $W_s^{jk}$ corresponding to the above equations for electromagnetic field are
\begin{eqnarray}
&&\epsilon _s=\gamma _s^2(\varepsilon _s+P_s{\bf v}_s^2),\quad{\bf S}_s=\gamma _s^2(\varepsilon _s+P_s){\bf v}_s, \nonumber\\
&&W_s^{ij}=\gamma _s^2(\varepsilon _s+P_s)v_s^jv_s^k+P_sg^{jk},\label{eq17}
\end{eqnarray}
where ${\bf v}_s$ is the fluid velocity, $n_s$ is the number density, $P_s$ is the pressure, and $\varepsilon _s$ is the total energy density defined by
\begin{equation}
\varepsilon _s=m_sn_s+P_s/(\gamma _g-1)\label{eq18}.
\end{equation}
The gas constant $\gamma _g$ take the value  $4/3$ for $T\rightarrow \infty $ and $5/3$ for $T\rightarrow 0$. Using the conservation of entropy the equation of state can be expressed by
\begin{equation}
\frac{D}{D\tau }\left(\frac{P_s}{n_s^{\gamma _g}}\right)=0.\label{eq19}
\end{equation}
The energy and momentum conservation Eqs. (\ref{eq14}) and (\ref{eq15}) coupling with each single perfect fluid of species $s$ to the electromagnetic field are
\begin{eqnarray}
&&\frac{1}{\alpha }\frac{\partial }{\partial t}P_s-\frac{1}{\alpha }\frac{\partial }{\partial t}[\gamma _s^2(\varepsilon _s+P_s)]-\nabla \cdot [\gamma _s^2(\varepsilon _s+P_s){\bf v}_s]\nonumber\\
&&+\gamma _sq_sn_s{\bf E}\cdot {\bf v}_s+2\gamma _s^2(\varepsilon _s+P_s){\bf a}\cdot {\bf v}_s=0,\label{eq20}\\
&&\gamma _s^2(\varepsilon _s+P_s)\left(\frac{1}{\alpha }\frac{\partial }{\partial t}+{\bf v}_s\cdot \nabla \right){\bf v}_s+\nabla P_s-\gamma _sq_sn_s({\bf E}\nonumber\\
&&+{\bf v}_s\times {\bf B})+{\bf v}_s\left(\gamma _sq_sn_s{\bf E}\cdot {\bf v}_s+\frac{1}{\alpha }\frac{\partial }{\partial t}P_s\right)\nonumber\\
&&+\gamma _s^2(\varepsilon _s+P_s)[{\bf v}_s({\bf v}_s\cdot {\bf a})-{\bf a}]=0\quad \label{eq21}.
\end{eqnarray}
Here, $\varepsilon _s$ is the internal energy density and $P_s$ is the fluid pressure. Similar energy and momentum conservation equations have previously been obtained by Buzzi et al \cite{twenty five, twenty six}. If, now, one sets $\alpha =1$ so that the acceleration goes to zero, these equations reduce to the corresponding special relativistic equations as given by SK \cite{four}. The curl and divergence operators in the 3+1 set of equations are covariant and can be derived in locally cartesian coordinates instead of the spherical three-metric using Rindler coordinate system, in which space is locally Cartesian, provides a good approximation to the Schwarzschild metric near the event horizon in the form
\begin{equation}
ds^2=-\alpha^2dt^2+dx^2+dy^2+dz^2, \label{eq22}
\end{equation}
where
\begin{equation}
x=r_H(\theta -\pi/2), y=r_H\phi, z=2r_H(1-\frac{2M}{r})^{-1/2}.\label{eq23}
\end{equation}
The standard lapse function in Rindler coordinates becomes $\alpha =z/2r_H$, where $r_H$ is the Schwarzschild radius. One of the advantages of the Rindler geometry is that it gives an example of the essential ideas of the horizon and the 3+1 split without the distracting complication of curved spatial three-metric.
The transformation from the FIDO comoving (fluid) frame to a coordinate frame of the metric given in Eq. (\ref{eq1}) involves a boost velocity, which is a simple
Lorenz boost with velocity $v_{\textrm{ff}}$ in the radial direction defined by
\begin{equation}
v_{\textrm{ff}}=(1-\alpha ^2)^{\frac{1}{2}}.\label{eq24}
\end{equation}
Then the relativistic Lorentz factor becomes
$\gamma_{\textrm{boost}}\equiv (1-v_{\textrm{ff}}^2)^{-1/2}=1/\alpha $.\\

{\bf \large 4. Wave Propagation in Radial Direction}\label {sec4}\\
Here we consider an incoming gravitational wave propagating in radial $z$ direction toward the event horizon in presence of an external static magnetic field ${\bf B}=B_0{\bf e}_{\hat z}$ and study gravitational waves excitation of small amplitude plasma waves, restricting our attention to a one-dimensional case. Introducing the following complex variables,
\begin{eqnarray}
v_s(z,t)=v_{sx}(z,t)+{\rm i}v_{sy}(z,t),\nonumber\\
B(z,t)=B_x(z,t)+{\rm i}B_y(z,t),\nonumber\\
E(z,t)=E_x(z,t)+{\rm i}E_y(z,t)\label{eq25},
\end{eqnarray}
the equation of continuity, Eq. (\ref{eq8}), and Poisson's Eq. (\ref{eq10}) for each fluid species can be written as
\begin{eqnarray}
\frac{\partial }{\partial t}(\gamma _sn_s)+\frac{\partial }{\partial z}(\alpha \gamma _sn_su_s)=0,\label{eq26}\\
\frac{\partial E_z}{\partial z}=4\pi (q_1n_1\gamma _1+q_2n_2\gamma _2).\label{eq27}
\end{eqnarray}
Two transverse parts obtaining from adding the ${\bf e}_{\hat y}$  component multiplied by $i$ to the ${\bf e}_{\hat x}$ component of the Maxwell\rq s Eqs. (\ref{eq11}) and (\ref{eq12}) may be written in single form as
\begin{eqnarray}
\left(\alpha ^2\frac{\partial ^2}{\partial z^2}+3\alpha \frac{\partial \alpha}{\partial z}\frac{\partial }{\partial z}-\frac{\partial ^2}{\partial t^2}+\left(\frac{\partial \alpha }{\partial z}\right)^2\right)E\nonumber\\
=4\pi e\alpha \frac{\partial }{\partial t}(n_2\gamma _2v_2-n_1\gamma _1v_1)\label{eq28}.
\end{eqnarray}
The transverse and longitudinal parts of the momentum conservation Eq. (\ref{eq21}) can be separated out of the form
\begin{eqnarray}
\rho _s\frac{Du_s}{D\tau }=q_sn_s\gamma _s\left(E_z+\frac{{\rm i}}{2}\left(v_sB^{\ast}-v^{\ast}_s B\right)\right)-\frac{\partial P_s}{\partial z}\nonumber\\
+(1-u^2_s)\rho _sa-u_s\left(q_sn_s\gamma _s{\bf E}\cdot {\bf v}_s+\frac{1}{\alpha }\frac{\partial P_s}{\partial t}\right),\label{eq29}\\
\rho _s\frac{Dv_s}{D\tau }=q_sn_s\gamma _s(E-{\rm i}v_sB_z+{\rm i}u_sB)\nonumber\\
-u_sv_s\rho _sa-v_s\left(q_sn_s\gamma _s{\bf E}\cdot {\bf v}_s+\frac{1}{\alpha }\frac{\partial P_s}{\partial t}\right),\label{eq30}
\end{eqnarray}
where the suffix star is the complex conjugate, $u_s$ the $z$ component of velocity, and the total energy density defined by $\rho _s=\gamma _s^2(\varepsilon _s+P_s)=\gamma _s^2(m_sn_s+\Gamma _gP_s)$ with $\Gamma _g=\gamma _g/(\gamma _g-1)$.
In order to investigate the transverse electromagnetic waves it is more convenient to work from a combination of Eq. (\ref{eq28}) and momentum conservation Eq. (\ref{eq29}). The longitudinal waves can be investigated by combining the longitudinal components of the continuity Eq. (\ref{eq26}), Poisson Eq. (\ref{eq27}), and the conservation of momentum Eq. (\ref{eq30}).\\

{\bf \hspace{.8 cm} \large 5. Linearized Equations}\label {sec5}\\
We linearize the two-fluid equations by considering a small perturbation. We introduce the quantities
\begin{eqnarray}
n_s(z,t)&=&n_{0s}(z)+\delta n_s(z,t)),\hspace{.1cm}v_s(z,t)=\delta v_s(z,t),\nonumber\\
P_s(z,t)&=&P_{0s}(z)+\delta P_s(z,t),\hspace{.1cm}{\bf E}(z,t)=\delta {\bf E}(z,t),\nonumber\\
u_s(z,t)&=&u_{0s}(z)+\delta u_s(z,t),\hspace{.1cm}{\bf B}(z,t)=\delta {\bf B}(z,t),\nonumber\\
\rho _s(z,t)&=&\rho _{0s}(z)+\delta \rho _s(z,t),\nonumber\\
{\bf B}_z(z,t)&=&{\bf B}_0(z)+\delta {\bf B}_z(z,t).\label{eq31}
\end{eqnarray}
The transverse set of two-fluid equations (\ref{eq28}) and (\ref{eq29}) are linearized using Eq. (\ref{eq31}) of the form
\begin{eqnarray}
\left(\alpha ^2\frac{\partial ^2}{\partial z^2}+3\alpha \frac{\partial \alpha }{\partial z}\frac{\partial }{\partial z}-\frac{\partial ^2}{\partial t^2}+\left(\frac{\partial \alpha }{\partial z}\right)^2\right)\delta E\nonumber\\
=4\pi e\alpha \left(n_{02}\gamma _{02}\frac{\partial \delta v_2}{\partial t}-n_{01}\gamma _{01}\frac{\partial \delta v_1}{\partial t}\right)\label{eq32},\\
\left(\alpha u_{0s}\frac{\partial }{\partial z}+\frac{\partial }{\partial t}-u_{0s}\frac{\partial \alpha }{\partial z}+\frac{{\rm i}\alpha q_s\gamma _{0s}n_{0s}B_0}{\rho _{0s}}\right)\frac{\partial \delta v_s}{\partial t}\nonumber\\
-\frac{\alpha q_s\gamma _{0s}n_{0s}}{\rho _{0s}}\left(\alpha u_{0s}\frac{\partial }{\partial z}+\frac{\partial }{\partial t}+u_{0s}\frac{\partial \alpha }{\partial z}\right)\delta E=0\label{eq33}.
\end{eqnarray}
The longitudinal set of equations (\ref{eq26}), (\ref{eq27}), and (\ref{eq30}) are linearized to obtain
\begin{eqnarray}
\gamma _{0s}\left(\frac{\partial }{\partial t}+u_{0s}\alpha \frac{\partial }{\partial z}+u_{0s}\frac{\partial \alpha }{\partial z}+\gamma _{0s}^2\alpha \frac{du_{0s}}{dz}\right)\delta n_s\nonumber\\
+\Bigg(\alpha \frac{\partial }{\partial z}+\frac{\partial \alpha }{\partial z}\Bigg)(n_{0s}\gamma _{0s}u_{0s})+n_{0s}\gamma _{0s}^3\Bigg[u_{0s}\frac{\partial }{\partial t}+\alpha \frac{\partial }{\partial z}\nonumber\\
+\frac{\partial \alpha }{\partial z}+\alpha \left(\frac{1}{n_{0s}}\frac{dn_{0s}}{dz}+3\gamma _{0s}^2u_{0s}\frac{du_{0s}}{dz}\right)\Bigg]\delta u_s=0\label{eq34},\\
\frac{\partial \delta E_z}{\partial z}=4\pi e(n_{02}\gamma _{02}-n_{01}\gamma _{01})+4\pi e(\gamma _{02}\delta n_2\nonumber\\
-\gamma _{01}\delta n_1)
+4\pi e(n_{02}u_{02}\gamma _{02}^3\delta u_2-n_{01}u_{01}\gamma _{01}^3\delta u_1),\quad\label{eq35}
\end{eqnarray}
and
\begin{eqnarray}
\left\{\frac{\partial}{\partial t}+u_{0s}\alpha \frac{\partial }{\partial z}+\gamma _{0s}^2\alpha (1+u^2_{0s}) \frac{du_{0s}}{dz}\right\}\delta u_s\nonumber\\
-\frac{\alpha q_s n_{0s}}{\rho_{0s}\gamma_{0s}}\delta E_z+\left(u_{0s}\alpha \frac{du_{0s}}{dz}+\frac{\alpha}{\rho_{0s}}\frac{dP_{0s}}{dz}+\frac{1}{\gamma_{0s}^2}\frac{\partial\alpha}{\partial z}\right)\nonumber\\
+\frac{1}{\gamma_{0s}^2 n_{0s}}\Bigg\{\frac{\gamma_{0s}^2 \gamma_g P_{0s}}{\rho_{0s}} \left(u_{0s}\frac{\partial}{\partial t}+\alpha\frac{\partial}{\partial z}\right)\nonumber\\
+\gamma_{0s}^2 \alpha\frac{\gamma_g P_{0s}}{\rho_{0s}}\left( \frac{1}{P_{0s}}\frac{dP_{0s}}{dz}-\frac{1}{n_{0s}}\frac{dn_{0s}}{dz}\right)\nonumber\\
+\left(1+\frac{\gamma_{0s}^2 \gamma_g P_{0s}}{\rho_{0s}}\right)\left(u_{0s}\gamma_{0s}^2 \alpha\frac{du_{0s}}{dz}+\frac{\partial\alpha}{\partial z}\right)\Bigg\}=0\label{eq36}.
\end{eqnarray}

{\bf \large 6. Dependence of the Unperturbed Values on $z$}\label {sec6}\\
Since the plasma is assumed to be falling in radial direction, the infall velocity can be defined as
\begin{equation}
u_{0s}(z)=v_{\rm ff}(z)=[1-\alpha ^2(z)]^{\frac{1}{2}}\label{eq37}.
\end{equation}
The unperturbed number density, pressure, temperature, and magnetic field can be determined directly from the equation of continuity. From Eq. (\ref{eq26}) it follows that $r^2\alpha \gamma _{0s}n_{0s}u_{0s}=\mbox{const.}=r_H^2\alpha _H\gamma _Hn_Hu_H$,
where the values with a subscript $H$ are the limiting values at the event horizon. The freefall velocity at the horizon becomes unity so that $u_H=1$. Since $u_{0s}=v_{\rm ff}$, $\gamma _{0s}=1/\alpha $; and hence $\alpha \gamma _{0s}=\alpha _H\gamma _H=1$. Also, because $v_{\rm ff}=(r_H/r)^\frac{1}{2}$, the number density, unperturbed pressure, temperature profile, and unperturbed magnetic field for each species can be written as follows:
\begin{eqnarray}
n_{0s}(z)=n_{Hs}v_{\rm ff}^3(z),\hspace{.2 cm}P_{0s}(z)=P_{Hs}v_{\rm ff}^{3\gamma _g}(z),\nonumber\\
T_{0s}=T_{Hs}v_{\rm ff}^{3(\gamma _g-1)}(z),),\hspace{.2 cm}B_0(z)=B_Hv_{\rm ff}^4(z),\label{eq38}
\end{eqnarray}
with $k_B=1$ and $P_{0s}=k_Bn_{0s}T_{0s}$.
The derivatives of the above quantities with respect to $z$ in Rindler coordinates expressed as
\begin{eqnarray}
\frac{dv_{\rm ff}}{dz}=\frac{du_{0s}}{dz}=-\frac{\alpha}{2r_H}\frac{1}{v_{\rm ff}}, \frac{dB_0}{dz}=-\frac{4\alpha}{2r_H}\frac{B_0}{v_{\rm ff}^2},\nonumber\\
\frac{dn_{0s}}{dz}=-\frac{3\alpha}{2r_H}\frac{n_{0s}}{v_{\rm ff}^2},\quad \frac{dP_{0s}}{dz}=-\frac{3\alpha}{2r_H}\frac{\gamma _gP_{0s}}{v_{\rm ff}^2}.\label{eq39}
\end{eqnarray}

{\bf \large 7. The WKB Approximation}\label {sec7}\\
We consider the infinitesimal displacements of waves with small amplitude toward the horizon in WKB approximation. We can write all the perturbations of the form
$f_0(z) \exp (i \int k(z) dz - i \omega t)$, where $f_0(z)$ and $k(z)$ are slowly varying function of $z$. Since the freefalling in any gravitational field occur only for locally, the correct WKB solution can be derived from the action principle developed by Heintzmann and Novello \cite{twenty four}. From this standpoint, the local dispersion relation and the instability can be anticipated from basic principle. The only scale in this problem is the black hole radius $r_H$, and the amplitude is small enough, we can ignore the internally reflated wave as long as $r_H/\lambda\gg 1$ because the amplitude of this wave vanishes as $e^{-r_H/\lambda}$, where  $\lambda = 2\pi/k(z)$. Thus, the set of transverse two-fluid equations, Eqs. (\ref{eq32}) and (\ref{eq33}) become
\begin{eqnarray}
&&\left({\alpha^2 k^2-\omega ^2}\right)\delta E\nonumber\\
&&={{\rm i}4\pi e\alpha \omega (n_{02}\gamma _{02}\delta v_2-n_{01}\gamma _{01}}\delta v_1),\label{eq40}\\
&&\omega \left(\alpha ku_{0s}-\omega +\frac{\alpha q_s\gamma _{0s}n_{0s}B_0}{\rho _{0s}}\right)\delta v_s\nonumber\\
&&-{\rm i}\alpha \frac{q_s\gamma _{0s}n_{0s}}{\rho _{0s}}\left(\alpha ku_{0s}-\omega \right)\delta E=0.\label{eq41}
\end{eqnarray}
Using Eq. (\ref{eq39}), the set of longitudinal two-fluid Eqs. (\ref{eq34})--(\ref{eq36}) when Fourier transformed become
\begin{eqnarray}
n_{0s}\gamma^2_{0s}\left(\alpha k-u_{0s}\omega\right)\delta u_s+\left(\alpha u_{0s}k-\omega\right)\delta n_s=0,\label{eq42}\\
\left(\alpha u_{0s}k-\omega\right)\delta u_s
+\frac{v^2_{Ts}}{2\gamma^2_{0s}n_{0s}}\left(\alpha k-u_{0s}\omega\right)\delta n_s\nonumber\\
+\frac{{\rm i}\alpha q_s n_{0s}}{\rho_{0s}\gamma_{0s}}\delta E_z=0,\label{eq43}\\
{\rm i}k\delta E_z=4\pi e(n_{02}\gamma _{02}-n_{01}\gamma _{01})+4\pi e(\gamma _{02}\delta n_2\nonumber\\
-\gamma _{01}\delta n_1)+4\pi e(n_{02}u_{02}\gamma _{02}^3\delta u_2-n_{01}u_{01}\gamma _{01}^3\delta u_1). \label{eq44}
\end{eqnarray}

{\bf \large 8. Transverse Electromagnetic Oscillations}\label {sec8}\\
The dispersion relation for transverse electromagnetic waves may be obtained from Eqs. (\ref{eq38}) and (\ref{eq39}) as
\begin{eqnarray}
k^2-\frac{\omega^2}{\alpha^2}=
\frac{\omega_{p1}^2\left(\frac{\omega}{
\alpha}-u_{01}k\right)}{u_{01}k-\frac{\omega}{\alpha}-\omega_{c1}}
+
\frac{\omega_{p2}^2\left(\frac{\omega}{
\alpha}-u_{02}k\right)}{u_{02}k-\frac{\omega}{\alpha}+\omega_{c2}},\label{eq45}
\end{eqnarray}
for either the electron-positron or electron-ion plasma.
The local plasma frequency $\omega _{ps}=\sqrt{{4\pi e^2\gamma _{0s}^2n_{0s}^2}/{\rho _{0s}}}$ depends on the local number density of each fluid species and the local cyclotron frequency

$\omega _{cs}={e^2\gamma _{0s}n_{0s}B_0}/{\rho_{0s}}$ depends upon  both the local number density and  magnetic field.

It is clear from the dispersion relations given in Eq. (\ref{eq45}) that the general relativistic effects enter only in the ratio $\omega/\alpha$. If one uses local time $d\tau = \alpha dt$ of the FIDO instead of a global coordinate time $t$, then these dispersion relations reduced to the special relativistic version as it should be according to the Einstein relativity principle.

If we are looking the background of the infall radial velocity of the fluid species toward the event horizon which is near unity closed to the event horizon (i.e., at $\alpha \rightarrow 0$ ) and decreases with the distance from it to zero (i.e., at $\alpha \rightarrow 1$ ) then the Eq. (\ref{eq45}) has simple analytic solutions at $\frac{\omega}{\alpha}\gg ku_{0s}$, $\frac{\omega}{\alpha}\approx ku_{0s}$, and $\frac{\omega}{\alpha}\ll ku_{0s}$. Let us consider these solutions.

{\bf 8.1 Electromagnetic waves with $\frac{\omega}{\alpha}\gg ku_{0s}$}\\
We considered the first case corresponds to the waves having phase velocity larger then the infall radial velocity (i.e., at $\frac{\omega}{\alpha k}\gg u_{0s}$).  In this case, the index of refraction can be approximated from Eq. (\ref{eq45}) of the form
\begin{eqnarray}
N^2_\alpha=\left(\frac{\alpha k}{\omega}\right)^2=1-\frac{\alpha^2}{\omega^2}\left[\frac{\omega_{p1}^2}{1+\frac{\omega_{c1}\alpha}{\omega}}+\frac{\omega_{p2}^2}{1-\frac{\omega_{c2}\alpha}{\omega}}\right]. \label{eq46}
\end{eqnarray}
This equation shows that this waves have a resonence not for electron, depends on the combine oscillations of electron with positron or ion. For electron-ion plasma, there is an asymmetry between the particle species due to the small mass ratio $m_e/m_i$, giving different order of magnitudes for the two local cyclotron frequencies, $\omega_{p1}$ and $\omega_{p2}$. Thus the resonances will occur vary different wave frequencies.

For electron-positron plasma ($m_e=m_i$, two species) Eq. (\ref{eq45}) can be written with $\omega_{c1}=\omega_{c2}$ and $\omega_{p1}=\omega_{p2}$ as
\begin{eqnarray}
N^2_\alpha=1-\frac{2\omega_{ps}^2}{\frac{\omega^2}{\alpha^2}-\omega_{cs}^2}.\label{eq47}
\end{eqnarray}
Which implies that two electromagnetic wave modes are exist: the upper branch represents the high frequency electromagnetic wave with frequency $\frac{\omega}{\alpha}>(2\omega_{ps}^2+\omega_{cs}^2)^{1/2}$ and another is the low frequency Alfv\'en wave with frequency $\frac{\omega}{\alpha}<\omega_{cs}$. On the other hand there appears a cut-off frequency $\frac{\omega}{\alpha}=(2\omega_{ps}^2+\omega_{cs}^2)^{1/2}$ in the range of electromagnetic wave. A resonance occurs when $\frac{\omega}{\alpha}=\omega_{ps}$. In this approximation our results agree with the results found by Daniel and Tajima \cite{ten}, and SK \cite{four} corresponding to general relativity and ultrarelativistic limit.\\

{\bf 8.2 Electromagnetic Waves with $\frac{\omega}{\alpha}\approx ku_{0s}$}\\
When the phase velocity of the waves approaching the fluid\rq s infall velocity, i.e., $\frac{\omega}{\alpha k}\approx u_{0s}$, we obtain from the dispersion relation given in Eq. (\ref{eq45}),
\begin{equation}
N^2_\alpha=\left(\frac{\alpha k}{\omega}\right)^2=1,\label{eq48}
\end{equation}
which shows that the electromagnetic waves reappear. Also, in this case we have $u^2_{0s}=1$, i.e., the infall radial velocity is maximum and is known as the freefall velocity of the fluid species.\\

{\bf 8.3 Electromagnetic Waves with $\frac{\omega}{\alpha}\ll ku_{0s}$}\\
Another solution can be simplified when the wave speed is far, or at least, is not very closed to the infall velocity (i.e., at $\frac{\omega}{\alpha k}\ll u_{0s}$) of the form
\begin{equation}
(u_{01}k-\omega_{c1})(u_{02}k-\omega_{c2})= -u_{01}u_{01}(\omega^2_{p1}+\omega^2_{p2}), \label{eq49}
\end{equation}
which for electron-positron plasma becomes
\begin{equation}
k^2u^2_{0s}=\omega^2_{cs}-2\omega_{ps}^2u^2_{0s}.\label{eq50}
\end{equation}
Since $\frac{\omega}{\alpha}\ll ku_{0s}<\omega_{cs}$, electromagnetic waves in a strong magnetized plasma reappear at frequency below the cyclotron frequency. When the plasma density is large and the field strength is small ($2\omega_{ps}^2u^2_{0s}>\omega^2_{cs}$) the wave number becomes imaginary ($k^2<0$) and the branch of a periodically damped and growing (i.e., unstable) oscillations exist. Similar instabilities of a relativistic plasma were found by Mikhailovskii \cite{thirty one}, and Zaslavskii and Moiseev \cite{thirty two}. Here the damping corresponds to $\textrm{Im}(k)>0 $ and growth to $\textrm{Im}(k)<0$.\\

{\bf \large 9. Longitudinal (Electrostatic) Oscillations}\label {sec9}\\
The longitudinal waves dispersion relation can be obtained from Eqs. (\ref{eq42}), (\ref{eq43}), and (\ref{eq44}) of the form
\begin{eqnarray}
1=\Bigg[\frac{\omega^2_{p1}}{(u_{01}k-\frac{\omega}{\alpha})^2-\frac{v^2_{T1}}{2}(k-\frac{\omega}{\alpha}u_{01})^2}\nonumber\\
+\frac{\omega^2_{p2}}{(u_{02}k-\frac{\omega}{\alpha})^2-\frac{v^2_{T2}}{2}(k-\frac{\omega}{\alpha}u_{02})^2}\Bigg],\label{eq51}
\end{eqnarray}
where $v^2_{Ts}=2\gamma_g \gamma^2_{0s}P_{0s}/\rho_{0s}$ is the fluid\rq s thermal velocity. Since the $\gamma^2_{0s}$ factor involved in the energy density $\rho_{0s}$ in the denominator, the $\gamma^2_{0s}$ factor in the numerator cancels out and therefore, the thermal velocity of the two-fluid plasmas are frame independent.

The general relativistic effects enter this dispersion relation like the transverse electromagnetic waves by the ratios of $\omega/\alpha$. If the lapse function is one, the effect of gravity vanishes and correspond to the special relativistic version.\\

{\bf 9.1 Longitudinal Waves with $\frac{\omega}{\alpha}\gg ku_{0s}$}\\
For $\frac{\omega}{\alpha}\gg ku_{0s}$, the dispersion relation given in Eq. (\ref{eq51}) can be approximated as
\begin{equation}
1=\Bigg[\frac{\omega^2_{p1}}{(\frac{\omega^2}{\alpha^2}-\frac{v^2_{T1}}{2}k^2)}+\frac{\omega^2_{p2}}{(\frac{\omega^2}{\alpha^2}-\frac{v^2_{T2}}{2}k^2)}\Bigg].\label{eq52}
\end{equation}
The low frequency mode corresponding to ion acoustic wave in the electron-ion plasma does not exist because the charge separation never occurs in that situation.
For electron-positron plasma, we can obtain the dispersion law of the waves under consideration from Eq. (\ref{eq52}) as
\begin{equation}
\frac{\omega^2}{\alpha^2}=2\omega^2_{ps}\left(1+\frac{\gamma_g}{2} \lambda^2_{Ds} k^2\right),\label{eq53}
\end{equation}
where $\lambda_{Ds}=\sqrt{k_B T_{os}/4\pi e^2 n_{0s}}$ is the the debye radius. Equation (\ref{eq53}) satisfies $\frac{\omega}{\alpha}\gg ku_{0s}$, if the second term in the right-hand side of this equation is small compared to the first term. This means that
\begin{equation}
\lambda^2_{Ds} k^2\ll 1,\label{eq53}
\end{equation}
which corresponds to rather long wavelength oscillations $k^2<<2\omega^2_{ps}$ at any arbitrary relativistic plasma temperature. In the limit of zero gravity (i.e., when $\alpha\rightarrow 1$) the above equation becomes $\omega^2=2\omega^2_{ps}(1+ \frac{\gamma_g}{2}\lambda^2_{Ds}k^2)$, which is the SK \cite{four} result with $\gamma_g=1$.\\

{\bf 9.2 Longitudinal Waves with $\frac{\omega}{\alpha} \approx ku_{0s}$}\\
In this approximation the dispersion relation given in Eq. (\ref{eq51}) reduces to
\begin{equation}
k^2=-\Bigg[\frac{\omega^2_{p1}}{(1-u^2_{01})\frac{v^2_{T1}}{2}}+\frac{\omega^2_{p2}}{(1-u^2_{02})\frac{v^2_{T2}}{2}}\Bigg],\label{eq54}
\end{equation}
It follows from Eq. (\ref{eq54}) with  $0<u_{0s}<1$ that in this case the branch of a periodically damped and growing (i.e., unstable) oscillations exist either for electron-positron or electron-ion plasma. When the plasma is freefalling onto the horizon Eq. (\ref{eq54}) becomes undefine, that is to say the isotropic pressure will not hold. For the electron-positron plasma Eq. (\ref{eq54}) can be written as
\begin{equation}
\frac{\omega^2}{\alpha^2}=\frac{2\omega^2_{ps}}{v^2_{T2}/2}\left(1+\frac{\gamma_g}{2} \lambda^2_{Ds} k^2\right),\label{eq53}
\end{equation}
Equation (\ref{eq53}) is valid for long wavelength oscillations and correesponding to the same dispersion curve given in Eq. (\ref{eq54}).

{\bf 9.3 Longitudinal Waves with $\frac{\omega}{\alpha}\ll ku_{0s}$}\\
For $\frac{\omega}{\alpha}\ll ku_{0s}$, we obtain from the Eq. (\ref{eq51}),
\begin{equation}
k^2=\Bigg[\frac{\omega^2_{p1}}{(u_{01}^2-\frac{v^2_{T1}}{2})}+\frac{\omega^2_{p2}}{(u_{02}^2-\frac{v^2_{T2}}{2})}\Bigg].\label{eq55}
\end{equation}
This equation shows that a singularity take place for each fluid species at the point for which the fluid\rq s velocity equals the half of the thermal velocity, $u^2_{0s}=\frac{1}{2}v^2_{Ts}$, i.e., the transonic radius begins to play a significant role for the longitudinal waves.  The position of transonic radius of each fluid mainly dependent on their limiting horizon temperature and determines the temperature at any given radius.  We have from Eq. (\ref{eq55}) for electron-positron plasma
\begin{equation}
\frac{\omega^2}{\alpha^2}\ll 2\omega^2_{ps}\left(1+\frac{\gamma_g}{2} \lambda^2_{Ds} k^2\right).\label{eq53}
\end{equation}
This equation is valid for the range of large wave numbers
\begin{equation}
\lambda^2_{Ds} k^2\gg 1.\label{eq53}
\end{equation}
Therefore, the formulae (\ref{eq55}) and (\ref{eq55}) can be qualitatively matched at $\lambda^2_{Ds} k^2\simeq 1$ and can be considered one and the same dispersion curve as various regions.\\

{\bf \qquad \large 10. Concluding Remarks}\label {sec10}\\
We have made an analytical study of transverse and longitudinal waves in an ideal relativistic two-fluid plasma, to advance some results obtained by different authors and to include corrections due to effects of general relativity.

Considering the transverse electromagnetic oscillations the dispersion curve are described by different analytical formulas. It follows from our analysis that one purely real Alfv\'en and high frequency electromagnetic modes are exist in high and low frequency limits. This results are is in agreement with the results of SK \cite{ten} for ultrarelativistic limit and of Daniel and Tajima \cite{ten} for general relativity. For very low or negligible frequency, damped and growing modes exist and the same conclusion also follows from the numerical calculations of Buzzi et al \cite{twenty five}.

For longitudinal oscillations, transonic radius begins to play a significant role. There exists only one real high frequency dispersion curve in high and low frequency limits as was found by SK in ultrarelativistic limit. For a very low frequency, damped and growing modes exist like the transverse electromagnetic waves. The presence of damped modes demonstrates that energy is being drained from the waves by the gravitational field and growing modes point out clearly that the  gravitational field is, in fact, feeding energy into the waves.

I am grateful to the anonymous referee for pointing out some mistakes and ambiguities in the first version of this manuscript and for giving me some references to improve the structure of this paper. I am glad to acknowledge the Editor Dr. Ronald C. Davision for sending me a paper related to this work.

\end{document}